\documentstyle[eqsecnum,preprint,aps]{revtex}
\tighten
\draft
\catcode`\@=11
\long\def\@makefntext#1{
\protect\noindent \hbox to 3.2pt {\hskip-.9pt
$^{{\ninerm\@thefnmark}}$\hfil}#1\hfill}		

\def\@makefnmark{\hbox to 0pt{$^{\@thefnmark}$\hss}}  

\def\ps@myheadings{\let\@mkboth\@gobbletwo
\def\@oddhead{\hbox{}
\rightmark\hfil\ninerm\thepage}
\def\@oddfoot{}\def\@evenhead{\ninerm\thepage\hfil
\leftmark\hbox{}}\def\@evenfoot{}
\def\sectionmark##1{}\def\subsectionmark##1{}}

\newcounter{sectionc}\newcounter{subsectionc}\newcounter{subsubsectionc}
\renewcommand{\section}[1] {\vspace*{0.6cm}\addtocounter{sectionc}{1}
\setcounter{subsectionc}{0}\setcounter{subsubsectionc}{0}\noindent
	{\normalsize\bf\thesectionc. #1}\par\vspace*{0.4cm}}
\renewcommand{\subsection}[1] {\vspace*{0.6cm}\addtocounter{subsectionc}{1}
	\setcounter{subsubsectionc}{0}\noindent
	{\normalsize\it\thesectionc.\thesubsectionc. #1}\par\vspace*{0.4cm}}
\renewcommand{\subsubsection}[1]
{\vspace*{0.6cm}\addtocounter{subsubsectionc}{1}
	\noindent {\normalsize\rm\thesectionc.\thesubsectionc.\thesubsubsectionc.
	#1}\par\vspace*{0.4cm}}

\newcounter{appendixc}
\newcounter{subappendixc}[appendixc]
\newcounter{subsubappendixc}[subappendixc]

\renewcommand{\appendix}[1] {\vspace*{0.6cm}
        \refstepcounter{appendixc}
        \setcounter{figure}{0}
        \setcounter{table}{0}
        \setcounter{equation}{0}
        \renewcommand{\thefigure}{\Alph{appendixc}.\arabic{figure}}
        \renewcommand{\thetable}{\Alph{appendixc}.\arabic{table}}
        \renewcommand{\theappendixc}{\Alph{appendixc}}
        \renewcommand{\theequation}{\Alph{appendixc}.\arabic{equation}}
        \noindent{\bf Appendix \theappendixc #1}\par\vspace*{0.4cm}}

\def\abstracts#1{{

\centering{\begin{minipage}{12.2truecm}\footnotesize\baselineskip=12pt\noindent
	\centerline{\footnotesize ABSTRACT}\vspace*{0.3cm}
	\parindent=0pt #1
	\end{minipage}}\par}}

\renewenvironment{thebibliography}[1]
	{\begin{list}{\arabic{enumi}.}
	{\usecounter{enumi}\setlength{\parsep}{0pt}
\setlength{\leftmargin 1.25cm}{\rightmargin 0pt}
	 \setlength{\itemsep}{0pt} \settowidth
	{\labelwidth}{#1.}\sloppy}}{\end{list}}

\newcounter{itemlistc}
\newcounter{romanlistc}
\newcounter{alphlistc}
\newcounter{arabiclistc}

\newcommand{\fcaption}[1]{
        \refstepcounter{figure}
        \setbox\@tempboxa = \hbox{\footnotesize Fig.~\thefigure. #1}
        \ifdim \wd\@tempboxa > 6in
           {\begin{center}
        \parbox{6in}{\footnotesize\baselineskip=12pt Fig.~\thefigure. #1}
            \end{center}}
        \else
             {\begin{center}
             {\footnotesize Fig.~\thefigure. #1}
              \end{center}}
        \fi}

\newcommand{\tcaption}[1]{
        \refstepcounter{table}
        \setbox\@tempboxa = \hbox{\footnotesize Table~\thetable. #1}
        \ifdim \wd\@tempboxa > 6in
           {\begin{center}
        \parbox{6in}{\footnotesize\baselineskip=12pt Table~\thetable. #1}
            \end{center}}
        \else
             {\begin{center}
             {\footnotesize Table~\thetable. #1}
              \end{center}}
        \fi}

\def\@citex[#1]#2{\if@filesw\immediate\write\@auxout
	{\string\citation{#2}}\fi
\def\@citea{}\@cite{\@for\@citeb:=#2\do
	{\@citea\def\@citea{,}\@ifundefined
	{b@\@citeb}{{\bf ?}\@warning
	{Citation `\@citeb' on page \thepage \space undefined}}
	{\csname b@\@citeb\endcsname}}}{#1}}

\newif\if@cghi
\def\cite{\@cghitrue\@ifnextchar [{\@tempswatrue
	\@citex}{\@tempswafalse\@citex[]}}
\def\citelow{\@cghifalse\@ifnextchar [{\@tempswatrue
	\@citex}{\@tempswafalse\@citex[]}}
\def\@cite#1#2{{$\null^{#1}$\if@tempswa\typeout
	{IJCGA warning: optional citation argument
	ignored: `#2'} \fi}}

 1
 1
 1

\font\ninerm=cmr9

\begin{document}
\begin{flushright}
 {CLNS 95/1381}
\end{flushright}
\bigskip
\centerline{\normalsize\bf SOME REMARKS ON SUPERSTRING PHENOMENOLOGY\footnote{
Talk presented at the International Symposium on Heavy Flavor and Electroweak
Theory, Beijing, China, 17-19 August 1995.}}
\baselineskip=22pt
\centerline{\footnotesize ZURAB KAKUSHADZE and S.-H. HENRY TYE}
\baselineskip=13pt
\centerline{\footnotesize\it Newman Laboratory of Nuclear Studies, Cornell
University}
\baselineskip=12pt
\centerline{\footnotesize\it Ithaca, NY 14853-5001, USA}
\centerline{\footnotesize E-mail: zurab@hepth.cornell.edu}
\centerline{\footnotesize and tye@hepth.cornell.edu}

\vspace*{0.9cm}
\abstracts{The present status of superstring phenomenology is briefly
discussed.}

\normalsize\baselineskip=15pt
\setcounter{footnote}{0}
\section{Introduction}

{}The number of consistent string models is huge and largely
unexplored. One should probably consider these models as different
classical vacua of a single theory, likely to be connected by
some continuous parameters, or moduli; if so, progress to locate the
correct vacuum that describes our universe will be extremely difficult,
mainly due to our lack of understanding of the complete moduli space
(or, the space of models), and its relevant underlying dynamics. Fortunately,
we know quite a lot about our universe. Assuming that superstring does
describe our universe, we may use the experimental data to guide
the search of the string model that governs our world. This
approach is known as superstring phenomenology.

{}It is reasonable to assume that nature has an underlying
supersymmetry (SUSY), so we may start with a specific heterotic string
model with $N=1$ space-time supersymmetry. Of course this supersymmetry
must be dynamically broken to reproduce the observed universe.
{\it A priori}, dynamical SUSY
breaking in string theory is very difficult to study, since it
involves the strong coupling regime of string theory. Fortunately, the
array of recent works~\cite{dual} on
string-string duality offers a plausible resolution to this problem.
String-string duality identifies the strong coupling limit of a
heterotic string model with the weak coupling limit of a specific
Type II string model. Since supersymmetry is broken, this Type II
model must have no space-time supersymmetry, {\it i.e.}, it is an $N=0$ model.
A typical string model will have an observable sector, which contains
the standard model, and a hidden sector (The need for a hidden
sector was recognized even before the modern advance of the superstring
theory~\cite{nilles}).
One then expects that dynamical supersymmetry breaking emerges from the
strongly coupled hidden sector of the heterotic string model, while
the observable sector remains weakly coupled.
This suggests that the
(semi-)classical description of the observable sector in the heterotic
string model may be quite close to what is really happening in nature,
while one can look at its Type II dual partner to study the supersymmetry
breaking pattern. This justifies the search for string models whose
visible sectors resemble nature. Such models should have hidden sectors
that contain asymptotically free gauge groups that become strongly
coupled at scales below the string scale, triggering SUSY breaking.
Fortunately, this seems to happen naturally in many examples.

{}What is a phenomenologically interesting string model? Roughly
speaking, we want an $N=1$ heterotic string model that has, in the visible
sector of its low energy effective (supergravity) field theory limit,
one of the following three features:\\
$\bullet$ ({\em i}) It contains the SUSY standard $SU(3)\otimes SU(2)
\otimes U(1)$ model. The simpliest standard model is the minimal
supersymmetric standard model (MSSM)~\cite{haber}. We shall generalize
the definition of MSSM to allow the inclusion of additional superheavy
particles. There are numerous examples of this type.\\
$\bullet$ ({\em ii}) It contains a SUSY grand unified theory (GUT).
So far, the only examples of this type have four chiral families.\\
$\bullet$ ({\em iii}) It contains a so-called "guided" grand unified model.
There are a number of examples of this type.

{}These possibilities are summarized in the following diagram:
\bigskip\bigskip
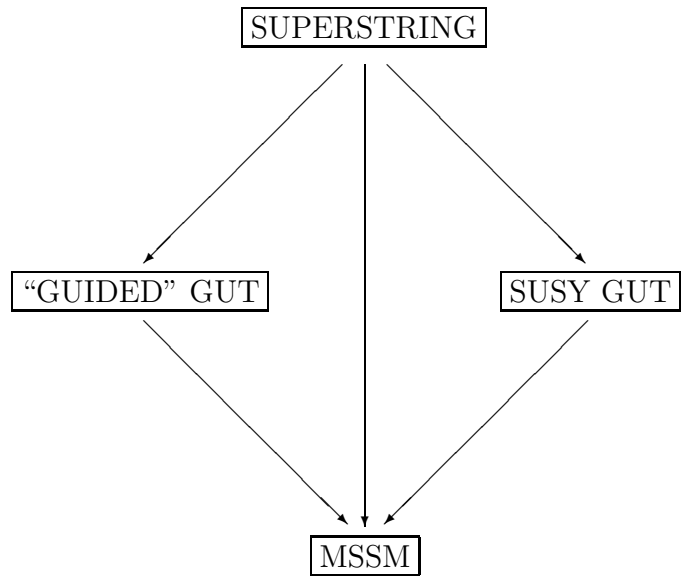
{
\begin{center}
\begin{picture}(200,200)(-55,-200)
\put(-3,0){\fbox{SUPERSTRING}}
\put(-90,-100){\fbox{``GUIDED'' GUT}}
\put(95,-100){\fbox{SUSY GUT}}
\put(35,-10){\vector(-1,-1){75}}
\put(52,-10){\vector(1,-1){75}}
\put(23,-200){\fbox{MSSM}}
\put(43.5,-10){\vector(0,-1){175}}
\put(-40,-107){\vector(1,-1){77}}
\put(128,-107){\vector(-1,-1){77}}
\end{picture}
\end{center}
\bigskip
Let us consider each of these cases in more detail.

\section{Standard Model}

{}Embedding the standard $SU(3)\otimes SU(2)\otimes U(1)$
model in string is relatively easy. There are many string models of this
type in the literature~\cite{stand}.

{}There are a couple of issues that are worth mentioning here.
At the string scale, the three gauge couplings of $SU(3)\otimes
SU(2)\otimes U(1)$ must be equal. Now, the string scale $M_{\mbox{string}}$ is
\begin{equation}
 M_{\mbox{string}}= g M_{\mbox{Planck}}/4\pi = 5 \times 10^{17} ~\mbox{GeV}~,
\end{equation}
where $g$ is the gauge coupling in a typical string model at the string scale.
In MSSM, the three gauge couplings unify at around
$3\times 10^{16}$ GeV~\cite{couplings}, which is definitely below the
string unification scale. Naively, this discrepancy is bothering.
Fortunately, in many string models that have been
looked at, this discrepancy is resolved by the
presence of string thresholds\cite{kap} and exotic matter fields\cite{dien},
which slightly modifies the running of the couplings in just the
right amount so that the coupling unification moves from
$3 \times 10^{16}$ GeV to
the string scale.

{}The total rank of the gauge groups in a generic heterotic string model
in four space-time dimensions is $22$ (for our counting purposes,
a $U(1)$ has rank $1$, and all solitons are massive) or less. If we
choose the standard model as the gauge group for
the visible sector, then the hidden (or semi-hidden) sector can
have a gauge group whose rank can be as large as $22-4=18$.
Such a large size allows for numerous possibilities.
Since the hidden sector interacts with the observable sector only via
the gravitational interaction, its impact on phenomenology is rather weak.
Furthermore, since the hidden sector is expected to involve non-perturbative
effects such as SUSY breaking, confrontation with
phenomenological constraints entails large theoretical uncertainties.
As a result, the possibilities of consistent choices for the hidden sector
are myriad and largely unexplored; and it is not that difficult to find
a string model that looks compatible with phenomenology, as testified
by many such claims in the literature. The difficulty is the lack of an
objective criteria that would select a particular model over all the others.
Finding the dual partner of any of these models would go a long way in
deciding its phenomenological viability, but this is not going to be easy.

\section{Grand Unified Theories}

{}Phenomenologically, grand unified theories (GUT) in the supergravity
framework (with three chiral families, in particular) are rather appealing.
These models have attracted a lot of attention. The LEP data on the
unification of the couplings~\cite{couplings} added further motivation
to this general idea. Recently, considerable effort has gone into the
analysis of the fermion mass matrices in supersymmetric grand unified models.
Generic features of these so called texture models~\cite{hr} look
very promising.
Furthermore they suggest that it is natural and desirable, and may even
be necessary, to embed such a framework in superstring.

{}So, how can we construct such a model? First, let us recall the
following facts. It is well known that in field theory adjoint Higgs
(or higher dimensional representations) is necessary for a grand unified
gauge group to break spontaneously to the $SU(3)\otimes SU(2)
\otimes U(1)$ gauge group of the standard model.
It is also known that, for Kac-Moody current algebras at level $1$, space-time
supersymmetry and chiral fermions cannot co-exist with massless
scalar fields in the adjoint (or higher dimensional) representations of
the gauge group in heterotic string models~\cite{dkv}.
 From these facts, we conclude that string GUT is possible only if the gauge
group is realized via Kac-Moody current algebras at levels higher than $1$.
Phenomenologically, level-$2$ or -$3$ gauge groups look promising.
Since higher-level current algebras have larger central charges,
the construction of such gauge groups reduces the size of the
hidden sector. This is an attractive feature because,
with smaller rank size for the hidden sector, the number of possibilities
is severely restricted.

{}The first higher level group in string theory appears in a
ten-dimensional heterotic string model. In the classification of
ten-dimensional models, all except one
have rank 16 gauge groups (the best known ones are the supersymmetric
$E_8 \otimes E_8$ and $SO(32)$ models).
The exception is non-supersymmetric and tachyonic, with a
(level-$2$) gauge group $E_8$~\cite{dh}.
Such uniqueness of higher
level gauge groups in ten dimensions disappears in lower space-time
dimensions. In four dimensions, in particular, there are models with
higher level gauge groups and $N=1$ space-time supersymmetry.
Suppose we start from this (unique) single $E_8$ model. Since it is
tachyonic, and so unstable, it must go to a stable string vacuum. (This is
similar to spontaneous symmetry breaking when we start from a Higgs potential
that, at zero vacuum expectation value, is tachyonic.)
Starting from the point in the moduli space that describes
this single $E_8$ model, it is possible that the nearby stable points
all describe level-2 gauge groups. So models with level-2 gauge groups
may be preferred.

{}The first string GUT was constructed by Lewellen~\cite{lew}.
In particular, he constructed an $SO(10)$ string GUT with four
chiral families. Next, Schwartz~\cite{schw} constructed an $SU(5)$ string
GUT, also with four chiral families. Both models are based on the
free fermionic string construction~\cite{klt}.
{\it A priori}, there is nothing that prevents one from
constructing a string GUT with three families.
However, string GUT models are quite complicated. The
construction of a typical higher-level string model involves a
non-abelian asymmetric orbifold~\cite{orb,nsv}. The rules for
model-building are easiest
to use in the free fermionic string model construction; so it is not surprising
that both of the known examples are constructed in this framework.
However, the fermionic string construction involves multiple
${\bf Z}_2$ twists, and so the number of families naturally comes out in
powers of $2$. To obtain three families, we may use the so-called
NAHE set~\cite{nahe}.
This is achieved with the following observation. One can always cut the
even number of families by half, in particular, one can cut two families
to one, since $2^0=1$; in the case of level-$1$ groups there is
plenty of room (i.e., $rank or central charge = 22$); so one can find
three different sets (or sectors) of matter fields, each with a single
family. As a result of this construction, these three families will have,
beyond the standard $SU(3)\otimes SU(2) \otimes U(1)$ quantum numbers,
different additional U(1) charges.
Unfortunately, attempts to incorporate the NAHE set with a higher-level
GUT gauge group have not been successful so far~\cite{lyk}. This is probably
because both the NAHE set and a higher-level gauge group need lots
of space, which is not available within the free fermionic construction.
Another promising possibility is have two families as a doublet, while
the third heavy family is by itself~\cite{cgl}. This way to incorporate
three families will presumably take up less space than the NAHE set and
is worth further investigation.

{}The above argument suggests that the free fermionic
string model-building is not suitable for string GUTs with three
families. But what if instead of a ${\bf Z}_2$ twist
we try a ${\bf Z}_3$ twist, which typically yields families in powers
of $3$. However, various attempts along this direction has been
unsuccessful so far. For example, using symmetric orbifolds,
$SO(10)$ models with adjoint or 54
Higgs can be constructed~\cite{afiu}, but they come with 4 families,
not 3. Also an $SU(5)$ model with 3
falimies was constructed~\cite{afiu}; unfortunately, the
model also has exotic chiral families.
Since higher-level gauge groups typically involve non-abelian
orbifolds, one is
led to consider asymmetric non-abelian orbifolds. Unfortunately,
any realistic, or semi-realistic, model involves a somewhat complicated
asymmetric non-abelian orbifold, and the rules for the construction of
such models are not streamlined enough to allow a straightforward search.
We believe this may be the reason that the search for string GUTs has
stalled at this moment.

{}As optimists, we may consider the failure in the search so far
as a good sign for string GUTs. In contrast to the standard model in
string theory, where there is a proliferation of possibilities, the
number of possible string GUTs should be very limited; it may even be
unique. In view of this possibility, we feel further efforts in the
search of string GUTs is worthwhile. As a first step, one should
streamline the existing rules for model-building in string theory.
Since the rules for the free fermionic
string model construction are rather simple, we decide to extend these
rules to include asymmetric~\cite{zurab} and non-abelian orbifolds.
In the following diagram the inside of the parallellogram
schematically indicates the subspace of string models that can
be obtained using free fermionic construction.

\bigskip
{\begin{center}
{\Large
\begin{center}
\begin{tabular}{|c|c|c|} \hline
 ORBIFOLD & SYMMETRIC & ASYMMETRIC \\ \hline
 ABELIAN & & \\ \hline
 NON-ABELIAN & & \\ \hline
\end{tabular}
\end{center}}
\begin{picture}(-105,0)(0,-24)
\put(30,15){\line(0,-1){30}}
\put(30,-15){\line(-1,0){60}}
\put(-30,-15){\line(0,1){30}}
\put(-30,15){\line(1,0){60}}
\end{picture}
\end{center}}
\bigskip

In the near future, we intend to use the rules for asymmetric
(non-)abelian model building to search for
phenomenologically interesting string models as well as their dual partners.

\section{``Guided'' GUTs}

{}Let us now turn to the third possibility. As pointed out earlier,
adjoint Higgs in a GUT comes only at the price of a higher-level
construction.
Actually, there is a way to have a grand unified gauge group without
adjoint Higgs. This possibility appears quite naturally in string models.
The first example of this type is the flipped
$SU(5)$ model~\cite{flipped}. It turns out that this is a special case of
a more general class of string models, which
we shall refer to as ``guided'' GUTs. Let the GUT gauge group be $G$
(typically $SU(5)$ or $SO(10)$). Then the gauge group of the guided GUT
is $G\otimes G^\prime$,
where we shall call the group $G^\prime$ the ``guiding'' group.
$G^\prime$ may
be bigger, equal, or smaller than $G$.
The basic idea of a ``guided'' GUT
involves the presence of Higgs fields $\phi$ which are in non-trivial
representations in both $G$ and $G^\prime$. When these Higgs fields
develop appropriate
vacuum expectation values that break $G\otimes G^\prime$ to $H$, where
$H\supseteq SU(3)\otimes SU(2)\otimes U(1)$, parts of $H$ comes
from both $G$ and $G^\prime$.

{}The Higgs fields employed in the ``guided'' GUTs
are in representations
lower than the adjoint representation. If $H= SU(3) \otimes SU(2)
\otimes U(1)_Y$, futher spontaneous symmetry breaking does
not require adjoint Higgs. Since no adjoint Higgs are needed
in these ``guided'' GUT models, level-$1$
gauge groups are perfectly adequate. Such gauge groups and their
respective Higgs fields appear quite frequently in string models. To see
if the Higgs fields do develop the appropriate vev, one has to study
the Higgs potential. In many instances, the appropriate vev lies in a
flat direction of the moduli space~\cite{font}, allowing for the
particular vev choice demanded from phenomenology. In fact, this is a
rather generic feature. The flipped $SU(5)$ string models have been
studied in great detail~\cite{flipped}. The $SU(5)^3$ GUT as well as an
$SO(10)^3$ GUT have been proposed~\cite{barb}.
The $SU(5)\otimes SU(5)$ as well as the $SO(10)^3$ string models
were recently constructed~\cite{fin},
For $G=SU(5)$, let us see how the Higgs mechanism works for the
following choices, $G^\prime =U(1), SU(5)$, or $SU(5)\otimes SU(5)$; in
each case, $H=SU(3)\otimes SU(2)\otimes U(1)_Y$.
The basic idea of the $SO(10)$ case is very similar.

{}The simpliest choice of $G^\prime =U(1)$ yields the flipped
$SU(5)$ model\cite{barr}.
It is instructive to start the discussion with $SO(10)$. Consider
$SO(10)\supset SU(5)\otimes U(1)_X$, then the spinor representation
of $SO(10)$ decomposes as ${\bf 16}={\bf 10}(1)
+ {\overline {\bf 5}}(-3) + {\bf 1}(5)$,
where the respective $U(1)_X$ charges $X$ are given in brackets.
Matter fields in the above combination will be anomaly-free.
For $SU(5)\supset SU(3)\otimes SU(2)\otimes U(1)_Z$, we have
\begin{eqnarray}
 &&{\bf 10}=({\bf 1},{\bf 1})(1) + ({\overline {\bf 3}},
   {\bf 1})(-2/3) +({\bf 3},{\bf 2})(1/6)~, \nonumber\\
 &&{\overline {\bf 5}}=({\overline {\bf 3}},{\bf 1})(1/3) + (
   {\bf 1},{\bf 2})(-1/2) ~,\nonumber\\
 &&{\bf 1} =({\bf 1},{\bf 1})(0)~,\nonumber
\end{eqnarray}
where the $U(1)_Z$ charges $Z$ are also explicitly shown. Let us start
from the guided GUT $SU(5)\otimes U(1)_X$ with three copies of chiral
fermions given above.
Suppose the component $(1,1)(1)(1)$ of a Higgs field ${\bf 10}(1)$ develops
a non-zero vacuum expectation value (vev). This field is singlet in
$SU(3)\otimes SU(2)$ and neutral in $U(1)_Y$, where the hypercharge $Y$ is a
linear combination of the charges $X$ and $Z$:
$Y/2 = (X-Z)/5$. So the
guided GUT spontaneously breaks to $SU(3)\otimes SU(2)\otimes U(1)_Y$,
the standard model gauge group.
In string theory, it is not hard to find a model with
three chiral families and this Higgs field. The
Higgs potential should develop a vev
along the $(1,1)(1)(1)$ direction.
As a minimum requirement, the Higgs potential
should be (perturbatively) flat along this direction.
These flipped $SU(5)$ string models have been extensively studied.
Since the rank of the gauge group of the hidden (plus semi-hidden)
sector of such string models can be as large as 17, the choices is
numerous.

{}In the case where $G^\prime=SU(5)$, the desired Higgs fields are
$\phi=({\bf 5},{\overline {\bf 5}})$ plus its complex conjugate.
String models of this type have been constructed~\cite{fin}.
The Higgs potential
either develops a vev at, or has a flat direction along,
$\langle\phi\rangle=
\langle\phi^*\rangle=\mbox{diag}(X,X,X,-3X/2,-3X/2)$.
With this vev, $SU(5)\otimes SU(5)$ spontaneously breaks to
$SU(3)\otimes SU(2)\otimes U(1)_Y$.

{}For $G^\prime =SU(5)\otimes SU(5)$, the Higgs fields are
$({\bf 1},{\bf 5},{\overline {\bf 5}})$,
$({\bf 5},{\overline {\bf 5}},{\bf 1})$ and
$({\overline {\bf 5}},{\bf 1},{\bf 5})$
plus their complex conjugates. These three Higgs fields develop vevs
of the form $\mbox{diag}(0,0,0,W,W)$,
$\mbox{diag}(X,X,X,Z,Z)$ and $\mbox{diag}(U,U,U,0,0)$
respectively.

{}Although guided GUTs are rather ugly from the field theory
point of view, they have a natural place in string models, due in
part to the presence of flat directions in the Higgs potential.
It is clearly worthwhile to explore these type of string models further.
So far, only flipped $SU(5)$ has been investigated in any detail.

\section{Summary}

{}Starting with a specific heterotic string model, one derives its low
energy effective (supergravity) field theory for energy scales below
the string scale; this approach allows one to use field theoretic
techniques to analyze many models and to impose phenomenological
constraints. This is the old approach.
With the recent understanding of string-string duality, one may be able
to work directly in string theory. In fact, for non-supersymmetric models,
we probably have a lot more control over its string version than over
its field theory version.
Hopefully, consistency of a dual pair
of string models will impose tighter constraints on the selection of
phenomenologically interesting string models.

\section{Acknowledgments}
This work was in part supported by the National Science Foundation.

%
%

\end{document}